# Optical design of a multi-resolution, single shot spectrograph

François Hénault[a], Florence Laurent[b]

[a] Institut de Planétologie et d'Astrophysique de Grenoble, Université Grenoble-Alpes, Centre National de la Recherche Scientifique, B.P. 53, 38041 Grenoble – France
[b] Univ Lyon, Univ Lyon1, Ens de Lyon, CNRS, Centre de Recherche Astrophysique de Lyon UMR5574, F-69230, Saint-Genis-Laval – France

**ABSTRACT**

Multi-object or integral field spectrographs are recognized techniques for achieving simultaneous spectroscopic observations of different or extended sky objects with a high multiplex factor. In this communication is described a complementary approach for realizing similar measurements under different spectral resolutions at the same time. We describe the basic principle of this new type of spectrometer, that is based on the utilization of an optical pupil slicer. An optical design inspired from an already studied instrument is then presented and commented for the sake of illustration. Technical issues about the pupil slicer and diffractive components are also discussed. We finally conclude on the potential advantages and drawbacks of the proposed system.

## 1 INTRODUCTION

### 1.1 A brief history of pupil slicing

In less than twenty years, image slicers became nearly standard equipment in the field of spectroscopy. They are now widely used in most astronomical observatories in order to detect and characterize distant galaxies or extra-solar planets. They are particularly appreciated for ground observations since they allow to acquire simultaneously all the spectra in all pixels of an astronomical image, therefore freezing the effects of atmospheric turbulence. They can usually be built from the design and assembly of fiber optics, micro-lens arrays, slicing and relaying mirrors, or the combination of any of the previous components.

So far, much less attention was paid to pupil slicers, who started being developed around the same years. It seems that their history begins at the multiple mirror telescope (MMT), an experimental facility composed of six co-phased telescopes of 1.8-m diameter each, acting together as a single monolithic telescope of 6.9-m diameter. But the six telescopes could also be coupled to a spectrograph in incoherent mode by stacking their individual images along its slit, where a set of six wedge plates were redirecting the beams towards its entrance pupil [1-2]. Here the pupil slicing function was ensured by the telescopes themselves.

In 1998, Iye *et al* published a short synthetic paper [3] where they described a specific device realizing the pupil slicing operation, based on an optical fiber arrangement as for the first generation of integral field spectrographs. The authors pointed out the main advantages offered by the technique, essentially improving the resolving power and throughput of a given spectrometer. However the most important advantage is probably the second (throughput enhancement), as illustrated by the recent integration of pupil slicing units inside two cross-dispersed echelle spectrographs currently under construction: ESPRESSO [4-5] where it is coupled with anamorphic optics, and SPIRou [6-7]. Here the throughput gain is achieved by generating and superimposing a number of sub-pupils onto the entrance pupil of a spectrograph of lower aperture than the input optical fibers.

One can finally note the existence of "sliced-pupil gratings" [8-9]. This is indeed an elegant solution to multiply the spectral resolution of a given transmissive diffraction grating (here a VPHG) by a factor of two or three, by means of Totally internal reflective (TIR) prisms increasing the angles of incidence on the VPHG: the TIR prisms were sliced into three smaller units to minimize the encumbrance of the dispersive element of the spectrometer. Here however pupil slicing as no real optical function and is rather used as a technological solution to mechanical design difficulties. We should also consider as pupil slicers all types of lenslet arrays integrated into Shack-Hartmann wavefront sensors for adaptive optics, making them the oldest ever realized.

## 1.2 Synthetic view

In this paper is described a new class of applications for pupil slicers. This approach ("Pupil slicing Type 2") is illustrated in Figure 1, comparing it with the now classical concept of image slicing, and with the most common use of pupil slicing in current spectrometers ("Type 1").

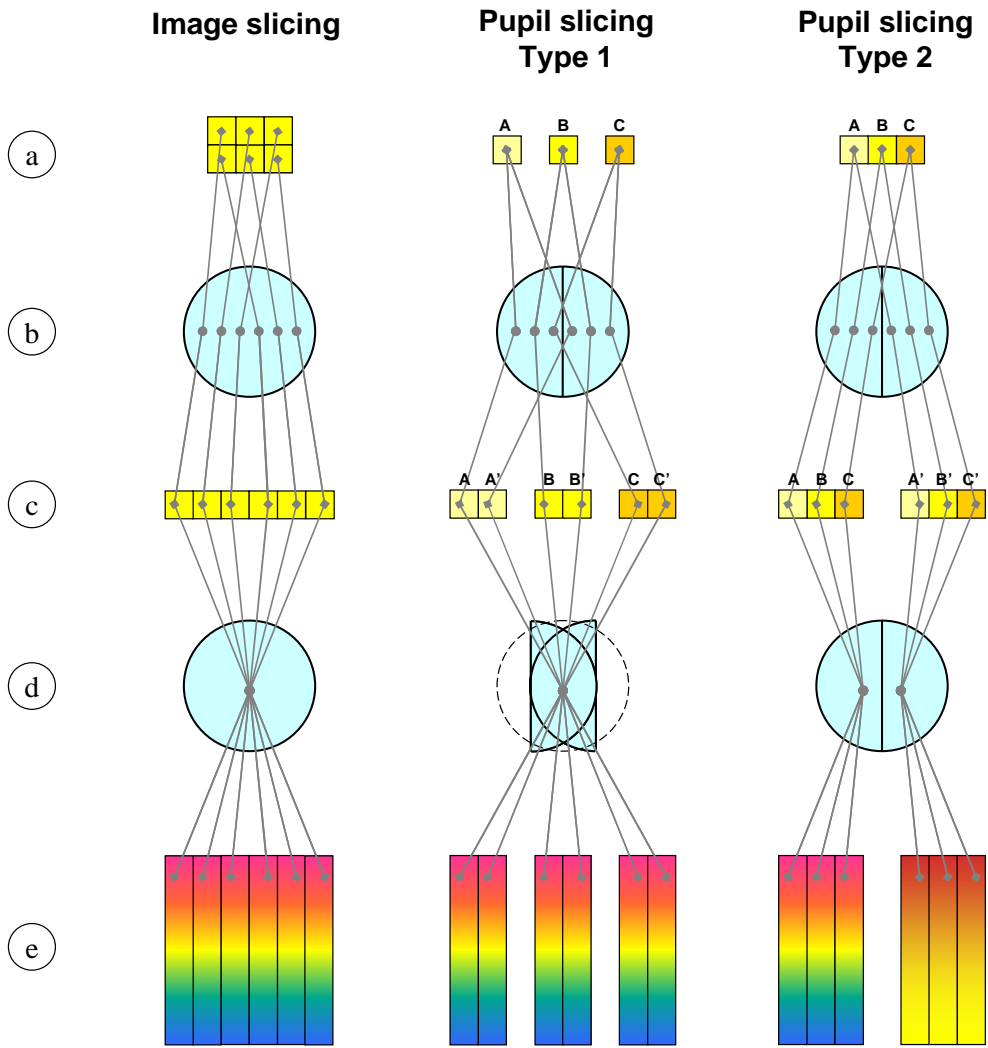

Figure 1: Schematic views of image slicing (left column), pupil slicing Type 1 (central column) and pupil slicing Type 2 (right column). From top to bottom the encircled letters respectively show: (a) Object in telescope focal plane. (b) Output pupil of slicing unit. (c) Object in spectrograph slit. (d) Input/output pupils of spectrograph. (e) Dispersed slit in spectrograph image plane. Grey lines are indicating the different optical pathways.

Image slicing

Image slicers are essentially used for integral field spectroscopy. Their general principle consists in slicing the images formed in telescope focal plane (a) into a number of small sections or "slices", then rearranging them side by side along the entrance slit of a spectrograph (c). The obtained pattern is often named "pseudo-slit". The telescope output pupil (b) is also reimaged at the input pupil of the spectrometer (d) – multiple intermediate pupil planes present inside the slicing unit are not shown on the figure. The mono-pupil spectrograph finally disperses the light perpendicularly to the pseudo-slit, thus allowing a simultaneous acquisition of the initial image and of the spectral decomposition of each image slice in the final image plane (e).

Pupil slicing Type 1

Nowadays pupil slicers are uniquely employed in association with spectrographs of the cross-dispersed echelle type, in order to characterizing extra-solar planets by the radial velocity method. Usually this kind of spectrometer can only support a very few input optical fibers (a), hence they are not imaging devices (for example the SPIRou instrument has only three input fibers: two for the observed star in two orthogonal polarization states and one calibration fiber). Those fibers noted A, B and C in Figure 1 are then reimaged by the pupil slicer (b) into a series of replicas stacked side by side along the entrance pseudo-slit of a spectrograph (c). A N-pupil slicer shall generate N replicas, each of typical intensity ratio 1/N (in the figure N = 2 and the replicas are noted A', B' and C'). Here the major functions of the slicing unit are twofold:

- Imaging the N pupil sections onto the spectrometer entrance pupil, and:
- Ensuring the best possible overlap of the beams on that plane (d).

The spectrograph will then act as a single-pupil instrument, though not uniformly illuminated. It further generates the spectral decomposition of the pseudo-slit on the detector plane (e). The final spectra of the input fibers A, B and C are computed by averaging them will their own replicas on the detector array (i.e. respectively AA', BB' and CC' in our case). At first glance, it then seems that only multi-object spectrographs or cross-dispersed echelle spectrometers could take benefit of a pupil slicer, because they can provide sufficient clearance between nearby spectra.

Pupil slicing Type 2

This new way of pupil slicing has a different purpose from Type 1, the main goals being to preserve the imaging capacity of the spectrograph and to provide the users with single-shot multi-resolution measurement ability. This can be achieved by setting different diffraction gratings in the spectrometer pupil plane (d), simultaneously illuminated by all objects located at the telescope focal plane. Those objects might be a number of optical fibers as in pupil slicing Type 1, but also slices of a rectangular image formed by the telescope (one may imagine an upstream image slicer transforming the rectangular image into a pseudo-slit).

The basic principle is illustrated on the right column of Figure 1. Either an image slice (named "mini-slit" in the following) or a row of optical fibers ABC located in the object plane (a) are replicated N times in the pseudo-slit plane of the spectrograph (c) by the N-slicing element (b). In the figure N = 2 and the replicated mini-slit is denoted A'B'C'. ABC and A'B'C' are separated by a gap of minimal length pre-defined by the optical designer. The slicing unit is also in charge of sub-pupils reimaging between the slicing plane (b) and the pupil of the spectrometer, where diffraction gratings of different resolving power are stacked side by side (d). Then the spectrograph acts as N instruments providing different spectral decompositions of the incoming light during the same time acquisition. Spatial and spectral data cubes can later be reconstructed using the same techniques as in integral field spectroscopy. It must be noted that pupil slicing of Type 2 could probably be applied to other instruments making use of various optical devices in their pupil plane, for example multiplex coronagraphs with different pupil masks.

In the following section is now presented a typical application of Pupil slicing Type 2, that is a multi-resolution single shot spectrometer inspired by an already existing instrument.

## 2 THE MULTI-RESOLUTION SPECTROGRAPH

### 2.1 Long slit and multi-resolution spectrographs

The ordinary method for realizing multiple observations of a given sky object at different spectral resolutions is illustrated by the upper panel of Figure 2. It simply consists in changing the dispersive elements of a classical long slit spectrometer by means of a dedicated mechanism. They can be prisms, diffraction gratings or prisms, often driven by translation stages or rotating wheels. However this procedure typically suffers from two fundamental drawbacks:

- Only one dispersive element can be utilized at the same time, hence the observations cannot be simultaneous. Meanwhile the atmospheric conditions may have changed, not only because of turbulence but also of atmospheric emission lines randomly appearing or vanishing in certain spectral bands.

- The mechanism is a critical issue by itself. It usually requires great effort in mechanical and electrical engineering to prevent potential failures. The problem is even worse when it has to operate in cryogenic environment (infrared or near IR instruments).

In opposition the multi-resolution spectrograph captures all the spectra simultaneously, then freezing the atmospheric perturbations and eliminating failure risks. Its major optical elements are schematically illustrated in the lower part of Figure 2, indicating the main coordinate frames (with the same encircled letters as in Figure 1). Following the sense of the incoming light, they are successively:

o The OXY plane of the entrance mini-slit (a), that can either be an image slice or a row of optical fibers aligned along the X-axis

o An optional cylindrical element compressing the beam along the Y and V axes, whose main function is to match the telescope pupil and the sub-pupils assembly (generally rectangular) in order to minimize flux losses. This component may also be useful to limiting geometrical aberrations inside the spectrometer.

o A pupil slicing element in the PUV plane (b), here depicted as a segmented mirror and reimaging the entrance mini-slit into a series of N replica In the pseudo-slit plane

o The O'X'Y' plane of the pseudo-slit (c). Near this plane but not shown on the figure should be located N converging elements used for pupil reimaging (either slit-mirrors or slit lenses)

o The collimating optics of the spectrograph

o N diffraction gratings located in the spectrometer pupil plane P'U'V' (d). Their dispersion direction is P'U', parallel to the X'-axis. The P'U'V' plane is assumed to be the actual aperture stop of the optical system, meanings that the beams are effectively limited by the edges of the diffraction gratings

o The camera optics of the spectrograph. Together with the collimator, it forms the image of the pseudo-slit plane onto the detector array

o The O"X"Y" plane of the detector array (e), where are formed N images of the entrance slit arranged side by side, each at different spectral resolutions

This short theoretical description can be considered as starting design rules for a pupil slicing spectrograph of Type 2. One may expect in particular that stringent requirements are set on imaging from planes OXY to O"X"Y", but also on pupil relaying between planes PUV and P'U'V'. In order to assess the feasibility of this type of spectrometers, we decided to realize a real case study, whose main specifications are defined in the following sub-section.

### 2.2 Requirements for a case study

In order to demonstrate the opportunity of a multiple spectral resolution spectrograph based on Pupil slicing Type 2, we preferred working on a concrete example coming from real-world astronomic and spectroscopic instrumentation. Our source of inspiration shall be the 4MOST instrument [10], currently under development for the VISTA 4-m telescope (Visible and infrared survey telescope for astronomy installed in Cerro Paranal, Chile). Here the basic idea is not to propose an alternative design for this instrument, which is already well engaged in the design phase, but to convince the reader of the feasibility of single shot multi-resolution spectrometer with the help of a didactic exercise of optical design.

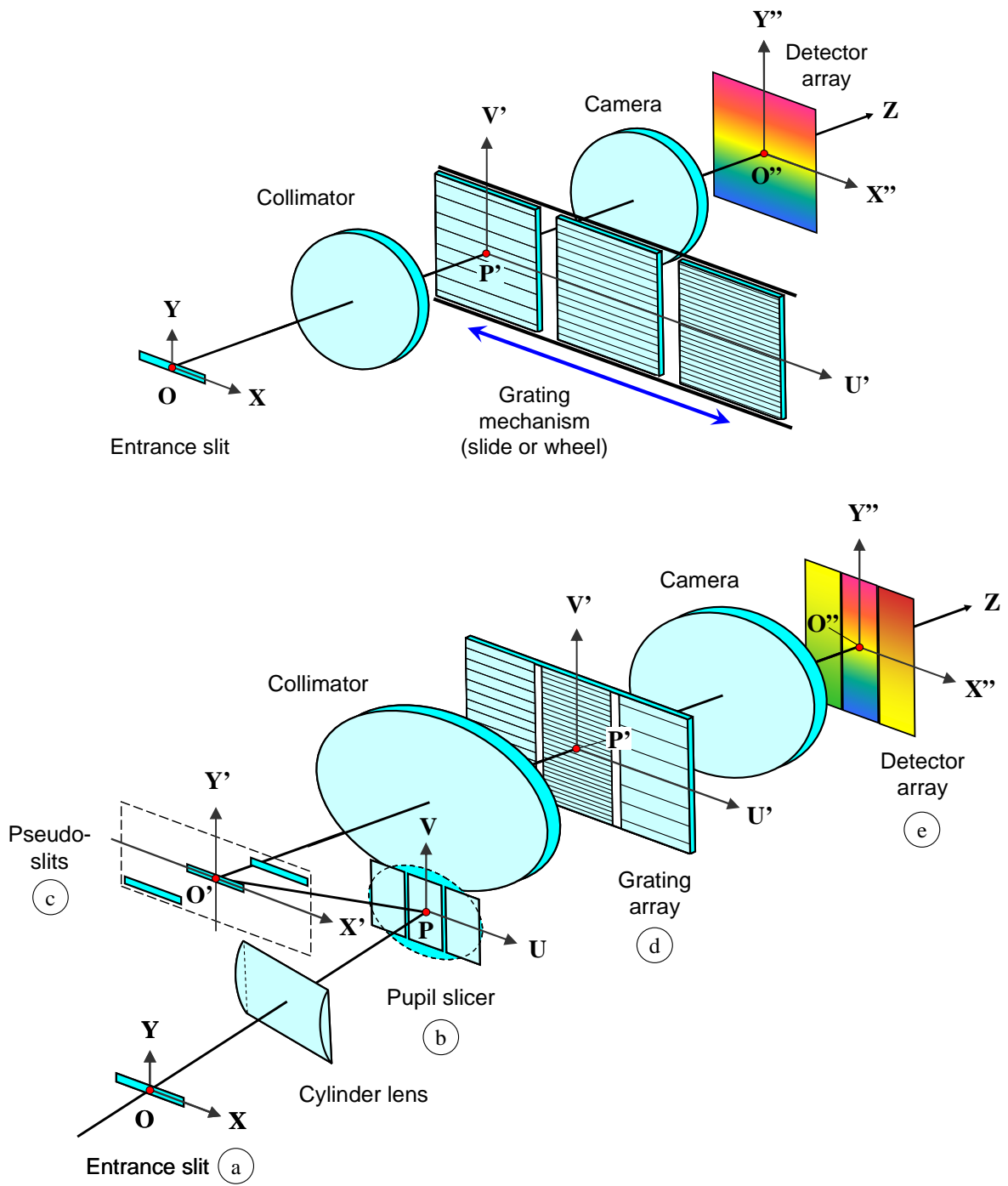

Figure 2: Schematic views of a long slit spectrometer (top panel) and of the proposed multi-resolution spectrograph (bottom panel with same encircled letters as in Figure 1).

The 4MOST instrument actually includes two three-channel low-resolution spectrograph (LRS) at a spectral resolution around 5000 covering a spectral band of 390-950 nm. It also comprises one three-channel high-resolution spectrograph (HRS) at a spectral resolution of 20000 and covering various spectral sub-bands. Those spectrometers are fed by a number of optical fibers aligned along their entrance slits (typically 800 fibers for the LRS). A complete description of the 4MOST instrument can be found in Ref. [11]. Our study is here limited to a single LRS channel that is 684-950 nm trying to append two higher-resolution channels (around 10000) operating simultaneously on sub-spectral bands 684-817 and 817-950 nm respectively. The main requirements of the original and modified spectrometers are summarized in Table 1 also including the characteristics of the employed detector array. As a general rule, we shall respect most of the elementary design choices of the original spectrograph, including the type and angle of diffraction grating, detector pixel number, and pixel size.

Table 1: Summary of 4MOST spectrometer requirements and their adaptation to the present study.

| SPECIFICATION | ORIGINAL REQUIREMENT | THIS WORK REQUIREMENT | Unit / Remark |
|---|---|---|---|
| **Low resolution spectrograph** | | | |
| Wavelength coverage | 684-950 | Config 1: 684-950<br>Config 2: 684-817<br>Config 3: 817-950 | nm<br>nm<br>nm |
| Spectral resolution | 5000 | Config 1: 5000<br>Configs 2 and 3: 10000 | on their full spectral range |
| Slit length | 159 | 60 | mm |
| Grating size | 180 x 197 | [3 x 80] x 160 | mm |
| Diffraction angle | 24.1 | See Table 3 | degrees |
| Groove frequency | 913 | See Table 3 | mm-1 |
| Camera aperture across slit | F/1.74 | F/2.2 | |
| **Detector array** | | | |
| Dimensions | 92.16 | 92.16 | mm |
| Number of pixels | 6144 | 6144 | |
| Pixel size | 15 | 15 | microns |

Transforming the LRS spectrograph of 4MOST into a multi-resolution instrument requires inserting pupil slicing optics in front of the spectrometer, and thus to define their preliminary requirements. The desired system can be simply divided into a pupil slicer subsystem and a spectrograph subsystem whose major requirements are given in Table 2. This very quick system analysis has essentially be driven by the geometrical characteristic of the existing LRS that exhibit two main difficulties: a very low image aperture number (F/D ≈ 1.75) and a quite long entrance slit (nearly 160 mm). With such parameters it is likely that both the numerical aperture and entrance slit length cannot be dramatically improved in order to cope with the unavoidable slit and pupil replications inherent to pupil slicing (with a mini-slit number N = 3 as selected for this study it would lead to F/D ≈ 0.58 and a pseudo-slit length of 480 mm, quite unrealistic values for optical designers). In order to keep more reasonable geometrical parameters, it was then chosen to:

- Restrict the entrance slit length of the system to around 60 mm so that it roughly matches the original LRS slit after replication and retro-magnification by the slicer optics. Assuming an inter-slit gap of 1-mm along the X'-axis in the intermediate image plane, it would still let 265 optical fibers from the real 4MOST instrument useable for multi-resolution spectroscopy.

- Reach a full numerical aperture F/D = 1.5 of the spectrograph camera along the U' and X'' axes, i.e. the direction along which the pupil of the spectrometer has to be divided into N = 3 sub-pupils. Hence the F/D number of each individual sub-pupil should be around 4.5, potentially hampering the spatial resolution of the spectrograph due to diffraction effects. It seems however that the resulting PSF enlargement should not be dominant when compared to the actual optical fiber diameter, pixel size, and image quality of the optics. It is then assumed that the spatial resolution is not affected significantly.

- Slightly relax the numerical aperture of the camera along the V' and Y'' axes in order to facilitate the optical design. The counterpart should be a slight decrease in spectral resolution that can be removed by reducing the diameter of the input optical fibers in the same proportion. For a F/D number of 2.2 across slit, the fiber diameter should be 68.2 µm in order to maintain the spectral sampling and resolution (instead of 85 µm for the original spectrograph).

Table 2: Main subsystems requirements of the multi-resolution spectrograph.

| Pupil slicer subsystem | | Spectrograph subsystem | |
|---|---|---|---|
| Telescope slit length | 60 mm | | |
| Telescope slit height | 68.2 microns | Collimator F/N X | 2 |
| Fiber diameter | 113.2 microns | Collimator F/N Y | 3 |
| Max fiber number | 265 | Camera F/N X | 1.5 |
| Telescope F/N | 3 | Camera F/N Y | 2.2 |
| Slicer F/N along X | 2 | Spectro Magnification X | 0.75 |
| Slicer F/N along Y | 3 | Spectro Magnification Y | 0.73 |
| Slicer magnification X | 0.67 | Pseudo-slit image length | 91.5 mm |
| Slicer magnification Y | 1.00 | Detector width | 92.16 mm |
| Mini-slit length | 40 mm | Pseudo-slit image height | 50.0 microns |
| Mini-slit heigth | 68.2 microns | Pixel size | 15 microns |
| Slit number | 3 | | |
| Inter-slit distance | 1 mm | | |
| Pseudo-slit length | 122 mm | | |

The preliminary subsystems requirements in Table 2 are now sufficient to start optical designing of the multi-resolution spectrograph with the help of a ray-tracing software. This is the scope of the next section.

## 3    OPTICAL DESIGN

In this section are presented the most important steps of the optical study, from initial definitions of the spectrograph (§ 3.1) and pupil slicing subsystem (§ 3.2) to their final integration and optimization into a single multi-resolution spectrometer assembly (§ 3.3). All computations were carried out using the Zemax™ ray-tracing program.

### 3.1    Spectrograph

As mentioned in the previous section, we start from the 684-950 nm spectral channel of the 4MOST LRS that is depicted on the left side of Figure 3. It comprises a curved entrance slit, a spherical collimating mirror, a spherical off-axis collimating mirror, an aspheric dioptric corrector, a volume phase holographic grating (VPHG) and a five-element focusing camera (lenses are noted L1 to L5 in the figure). The image quality of the system is illustrated on the right side of the figure. It shows a map of the root mean square (RMS) radius of images formed by the spectrometer at various locations in the O"X"Y" plane of the detector array, for different field positions (9 along X"-axis) and wavelengths (7 along Y"-axis). When averaged on the whole O"X"Y" plane the mean RMS spot radius is found to be close to 7 µm. This excellent performance is probably out of the capacity of a multiple resolution spectrograph, thus a less ambitious target was selected for our study, that is achieving an averaged RMS spot radius lower than 30 µm (i.e. two camera pixels). For that purpose, the first step of our optical design exercise consists in preparing the LRS for future integration into the multi-resolution instrument with the following goals:

- To match all sub-system requirements on the right side of Table 2, especially transforming the circular pupil into an elliptical exit pupil with aperture numbers F/D = 1.5 along the X"-axis and 2.2 along the Y"-axis,
- To preserve the locations of the real pupil of the spectrometer (actually the VPHG) and of its entrance pupil.
- To maintain a 45 degrees. angular deviation between the collimator and camera optical axes imposed by the VPHG characteristics.
- To respect the initial choice of optical materials for lenses L1 to L5.
- To not degrade image quality in order to keep sufficient margin for the next design steps.

Achieving these objectives will be made easier by means of a few simplifications of the original design. The most appreciable one comes from the reduced spectral band of 684-950 nm allowing significant image quality improvement

(originally the LRS cameras are made of the same lenses for the three different spectral channels, hence they must fulfill the performance requirements on the full 390-950 nm range). Also, the dichroic plate and the convex slit let lens at the entrance of the spectrograph can be removed from the original optical layout.

The result of the optimization is illustrated in Figure 4. The general appearance of the spectrometer (shown on the left side of the figure) remains the same, though a few simplifications could be made: two surface aspherizations could be removed on the off-axis corrector and L4 lens respectively. The mean RMS spot radius in the image plane (right side of the figure) has significantly improved following reduction of the spectral bandwidth and is now below 7 µm. This is as good as the original spectrograph and may provide us with sufficient margin for the following steps of the study.

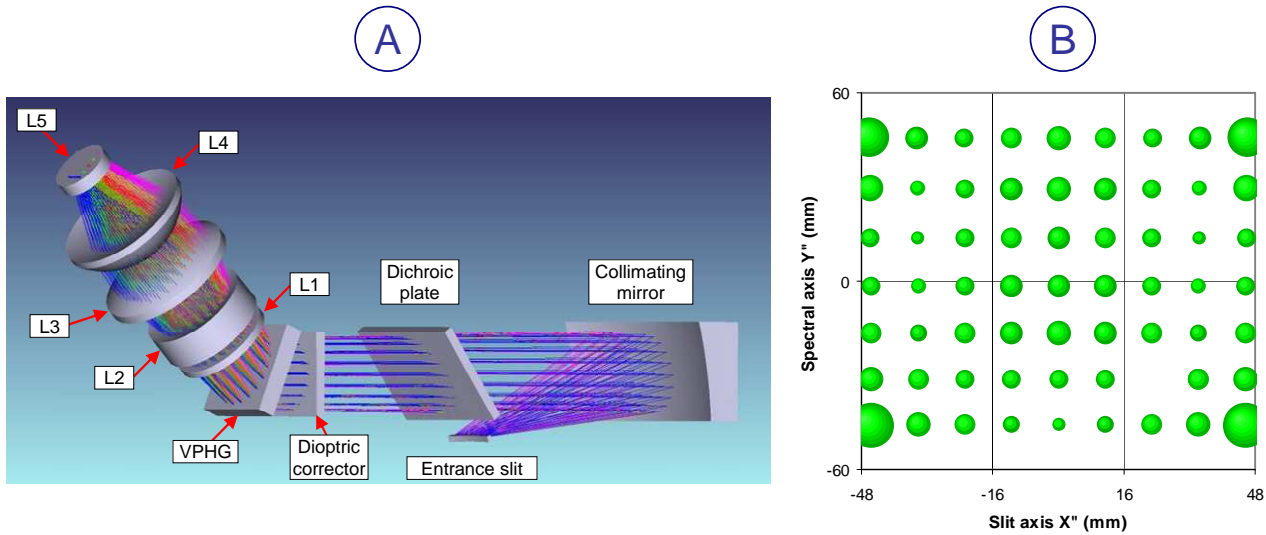

Figure 3: The original low-resolution spectrograph of 4MOST: (A) Shaded model of the spectrograph. (B) Image map in the O"X"Y" plane, showing the RMS spot radius on the detector array for various field positions and wavelengths (short or "blue" wavelengths on the -Y" side, long or "red" wavelengths on the +Y" side).

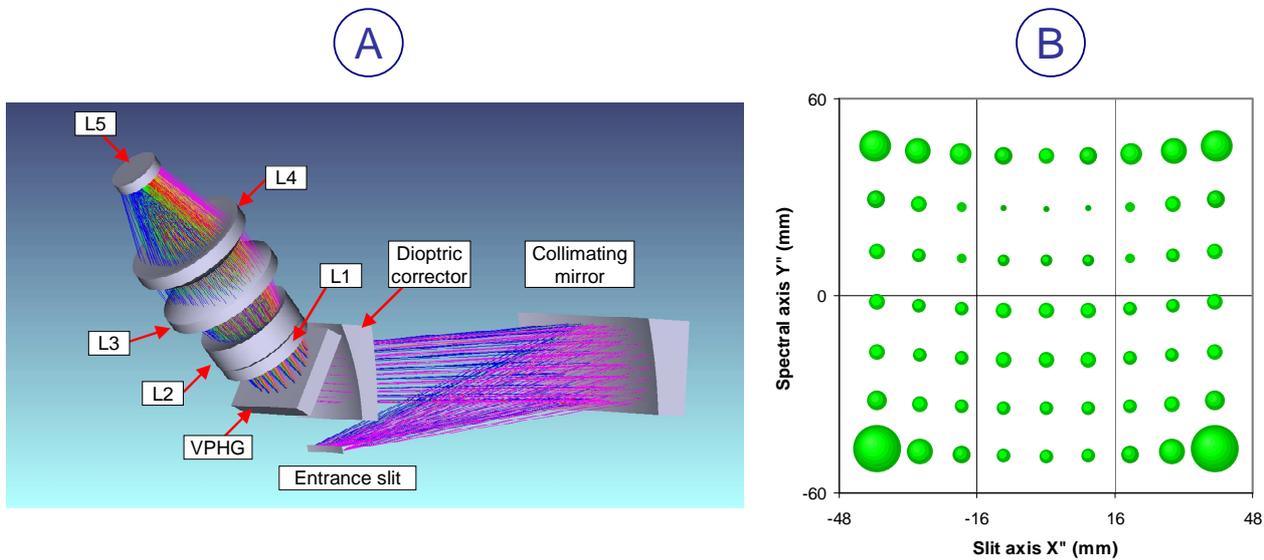

Figure 4: Modified LRS pre-optimized for multiple resolution operation. Same type of illustrations as in Figure 3.

## 3.2 Pupil slicer

The preliminary requirements of the image slicing sub-system are summarized on the left side of Table 2. As mentioned in sub- section 2.2 this kind of optics are essentially composed of two major elements, the first one actually slicing the pupil of the system into a number of sup-pupils and the second one, located near the pseudo-slit plane, reimaging them onto the pupil of the spectrograph. Moreover, those elements usually are of the reflective type; i.e. tilted or decentred spherical mirrors: this general trend is probably inherited from image slicing systems, where refractive optical elements are indeed extremely uncommon. Owing to some stringent requirements in Table 2, however (long pseudo-slit, low aperture numbers especially along X'-axis), we finally adopted that last solution considering that:

- Image slicers of this kind were already manufactured and successfully tested in the laboratory for the first version of the image slicing sub-system of the MUSE instrument [12-14].

- They provide more flexibility in optical designing due to a larger number of free parameters (lenses curvatures, thicknesses and choice of the optical material).

- Transmission losses by anti-reflection coated lenses are not significantly worse than mirror reflection, especially at non-normal incidence angles.

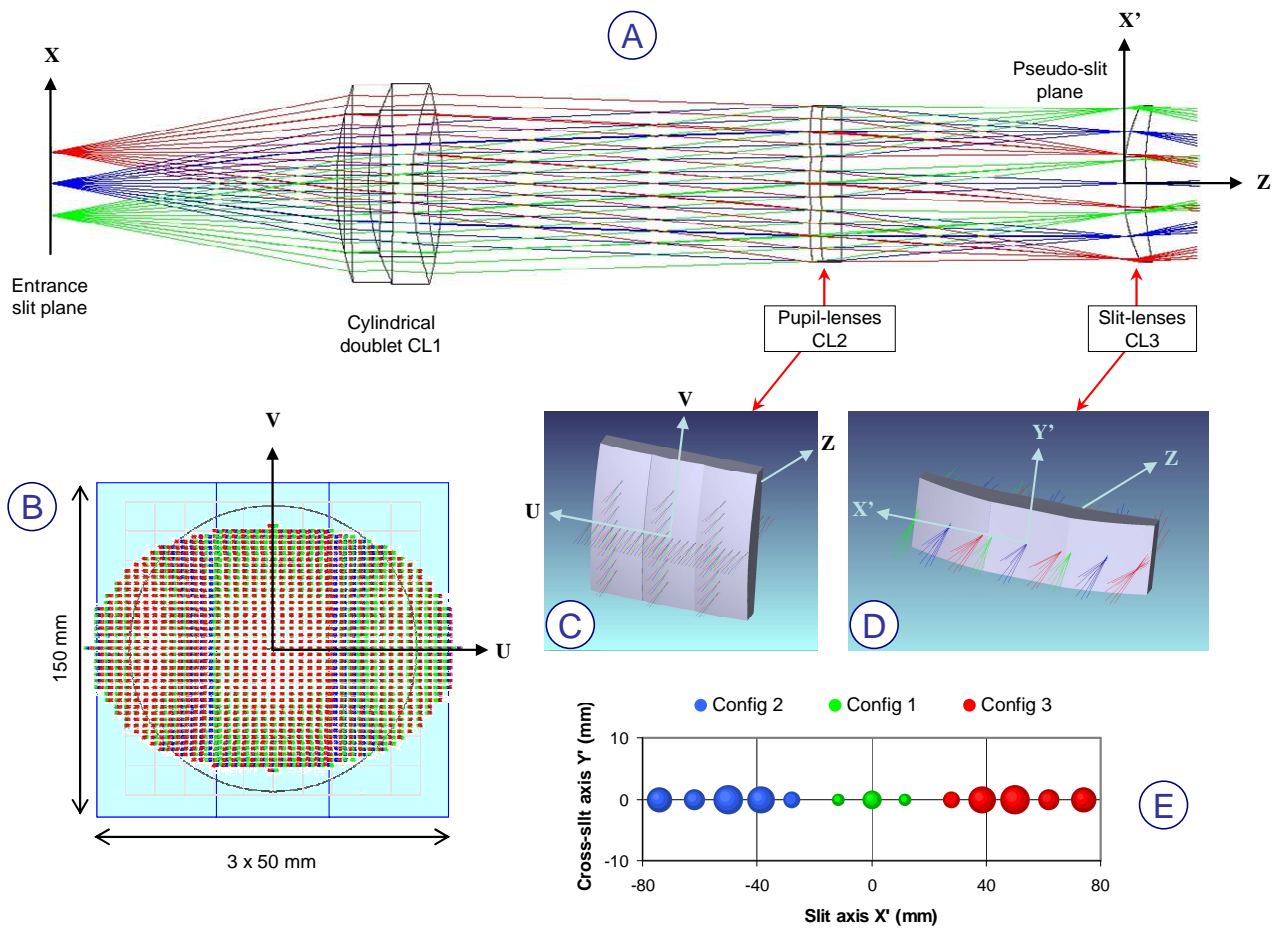

Figure 5: Preliminary pupil slicing sub-system. (A) General optical layout. (B) Pupil slicing in the PUV plane, here the entrance plane of CL2. (C) 3-D view of slicing pupil-lenses CL2. (D) 3-D view of slit-lenses CL3. (E) Diagram showing the RMS spot radii along the pseudo-slit axis X'.

The preliminary definition of the pupil slicing sub-system is illustrated in Figure 5. The general optical layout is shown on the upper part of the figure and is composed of:

- An achromatic "cylindrical" doublet noted CL1 compressing the pupil height along the V-axis in order to achieve the required anamorphic ratio of 2/3. CL1 is actually made of two toroïdal lenses.
- The pupil slicing element (shown on panel C), hereafter named "pupil-lenses" and noted CL2. It is a set of three toroïdal lenses stacked side by side along the V-axis, cutting the elliptic pupil into three sections and forming three images of the entrance slit along the X'-axis.
- The "slit-lenses" element noted CL3 and shown on panel D[1]. Its main function is to conjugate optically the pupil-lenses with the entrance pupil of the spectrometer described in sub-section 3.1.

Figure 5 also shows the optical footprint of the beam in the slicing plane PUV (panel B) and the way it is split into the three different sub-pupils. The optical model makes use of the multi-configuration mode of Zemax, each configuration being associated with a given sub-pupil: configurations n° 1, 2 and 3 respectively correspond to the central, -U and +U pupil sections, and thus the central, -X' and +X' mini-slits. The achieved RMS spot radii along the pseudo-slit axis are illustrated in the panel E of the figure. Their averaged value is currently around 200 µm. The optimization of the pupil slicing subsystem was not pushed further, knowing that it should be significantly modified when coupling it with the spectrograph. This story is told in the next sub-section.

### 3.3 Slicer and spectrograph assembly

Coupling the pupil slicing and spectrograph sub-systems pre-optimized in the previous sub-sections was not a straightforward task. In particular, three issues required special attention:

1) Although the pseudo-slit length along the X'-axis and the numerical apertures of the pupil slicer and spectrometer were carefully matched during the sub-systems design (following the prescription of Table 2), a strong pupil aberration appeared between the PUV and P'U'V' planes when they were assembled. Because this aberration essentially originates the spherical off-axis collimating mirror, the latter was replaced with an additional cylindrical doublet noted CL4. Incidentally, a flat folding mirror has also been added to keep the same general aspect to the spectrograph (see the upper panel of Figure 6).

2) In order to meet the specifications of Table 1, different types of dispersive elements have to be inserted in the three sub-pupils of the spectrometer. An unexpected difficulty arises because of the imposed angular deviation of 45 degs. between the incident and diffracted beams on the VPHG. Respecting this condition at a given central wavelength while attaining a given spectral resolution (here 10000) is a well constrained problem that can be solved by replacing the exit plane and parallel glass plate of the VPHG in configuration 1 by a refractive prism in configurations 2 and 3. The choice of the diffraction gratings and prisms parameters for the three different configurations is given in Table 3 (for configuration 1 they are the same as in § 3.1). Together they form the Dispersive elements assembly (DEA) depicted in Figure 6-D. However the presence of the prisms generates strong differential chromatism between the three configurations, which must be compensated for by the sole non-common optical elements.

Table 3: Main specifications of dispersive elements.

|  | Configuration 1 | Configuration 2 | Configuration 3 |
|---:|:---:|:---:|:---:|
| Spectral domain (nm) | 684-950 | 684-817 | 817-950 |
| Grating groove frequency ($\mu m^{-1}$) | 0.913 | 1.620 | 1.497 |
| Prism material | BK7 | SF11 | SF11 |
| Prism angle (degrees) | 0 | 37.16 | 46.55 |
| Prism base (mm) | 80 | 140 | 165 |

---

[1] It is in principle possible to replace the slit-lenses with a single optical element conjugating the PUV and P'U'V' planes. Such a lens with bi-conic shape was actually designed at this stage of the study, achieving same performance as the slit-lenses. However this option was discarded during the slicer and spectrograph assembly due to differential chromatism generated by the prisms in the three different configurations (see § 3.3).

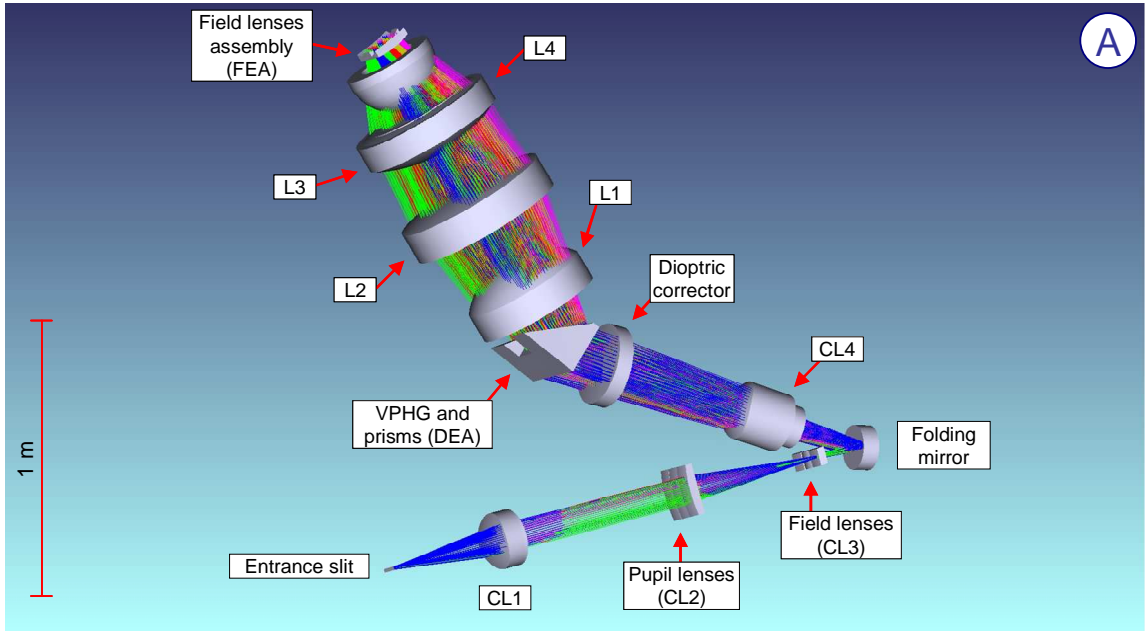
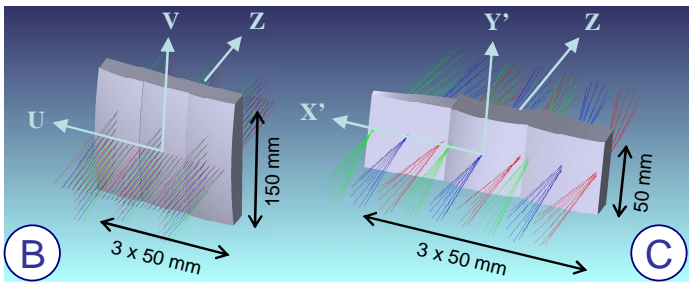
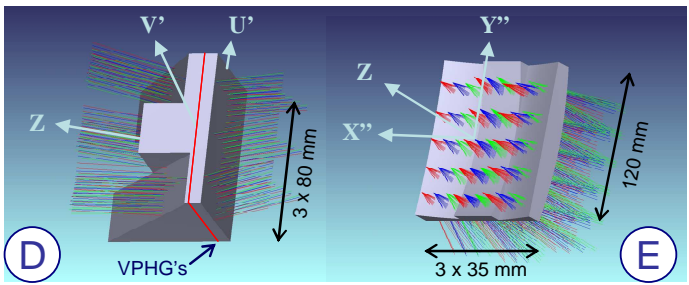
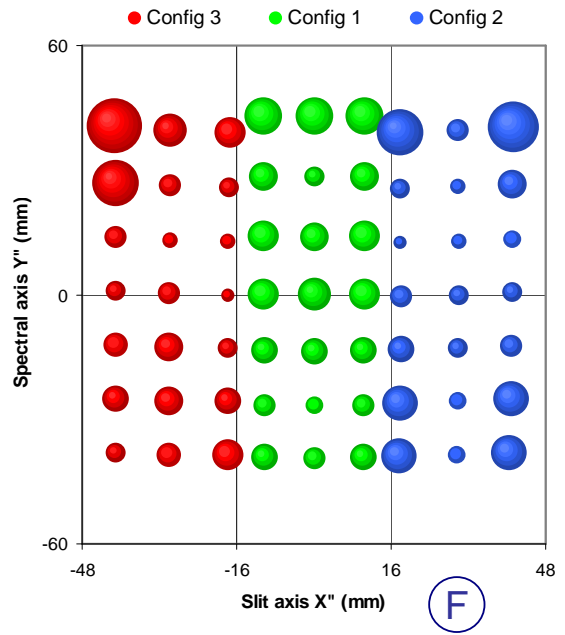

Figure 6: The final multi-resolution spectrograph. (A) General optical layout. (B) 3-D view of slicing pupil-lenses CL2. (C) 3-D view of slit-lenses CL3. (D) 3-D view of the dispersive elements assembly. (E) 3-D view of the field lenses assembly. (F) Diagram of the RMS spot radii in the image plane for various field positions and wavelengths (short or "blue" wavelengths on the -Y" side, long or "red" wavelengths on the +Y" side).

3) In order to mitigate this chromatism, it was necessary to split the field lens L5 of the spectrograph into three different lens sections, one for each configuration. The optical design after optimization of the system results in a Field lenses assembly (FLA) that is shown in Figure 6-E.

After assembling the pupil slicing and spectrograph sub-systems and optimizing them according to the previous guidelines, we finally arrived to a solution illustrated in Figure 6. It shows the general optical layout (A) and three-dimensional views of the pupil-lenses (B), slit-lenses (C), DEA (D) and FLA (E). It can be noted that the slit lenses (C) evolved significantly with respect to their pre-design (a dissymmetry between configurations 2 and 3 is clearly visible in Figure 5-D), while changes in the pupil slicing lenses (Figure 5-C and 6-B) are hardly perceptible. The achieved performance in terms of RMS spot radii on the detector array is also illustrated in Figure 6-F. The averaged RMS spot radius is found to be 26 µm, a number that is well within our initial goal of 30 µm and is very promising in view of future re-optimizing on the basis of real scientific and technical specifications. We then consider the feasibility of the multiple resolution single shot spectrograph as validated. Such re-optimization should aim at improving the image quality but also spectral and spatial resolutions, and minimizing the inter-slit distances at the image plane without generating crosstalk.

## 4   CONCLUSION

In this paper was presented the principle of a single shot multiple resolution spectrograph enabling to freeze the effects of atmospheric perturbations and including no mechanism. The concept makes use of a new way of realizing pupil slicing, named "Type 2" in the manuscript. The basic optical layout was described, and a preliminary system analysis has been carried out. We also provided an example of optical design inspired by a real-world astronomical spectrograph (the 4MOST instrument). Equivalent image quality performance was finally achieved. It must be pointed out that the spectrographs of 4MOST actually are very demanding systems: with an entrance slit of 160 mm, an image F-number of 1.75 and a 6K x 6K detector array, they are carrying an enormous optical etendue, making them particularly difficult to design, manufacture and align. Their successful adaption to the theoretical case of a multiple resolution spectrometer with comparable science requirements and performance (even at the price of a few additional conic or toroïdal surfaces) is certainly a strong argument for demonstrating the feasibility of this type of instrument in the future.

FH acknowledges funding help from the French "Action spécifique haute résolution angulaire" (ASHRA) managed by CNRS-INSU. He also thank his colleagues G. Duvert and P. Rabou for their advices and careful reading of the manuscript.